\documentclass[twocolumn]{aastex631}

\begin{document}

\title{Constraints on the X-ray and Very High Energy $\gamma$-ray Flux from Supernova Remnant W44}



\author{A.~Archer}\affiliation{Department of Physics and Astronomy, DePauw University, Greencastle, IN 46135-0037, USA}

\author[0000-0002-3886-3739]{P.~Bangale}\affiliation{Department of Physics and Astronomy and the Bartol Research Institute, University of Delaware, Newark, DE 19716, USA}

\author[0000-0002-9675-7328]{J.~T.~Bartkoske}\affiliation{Department of Physics and Astronomy, University of Utah, Salt Lake City, UT 84112, USA}

\author[0000-0003-2098-170X]{W.~Benbow}\affiliation{Center for Astrophysics $|$ Harvard \& Smithsonian, Cambridge, MA 02138, USA}

\author[0000-0001-6391-9661]{J.~H.~Buckley}\affiliation{Department of Physics, Washington University, St. Louis, MO 63130, USA}

\author[0009-0001-5719-936X]{Y.~Chen}\affiliation{Department of Physics and Astronomy, University of California, Los Angeles, CA 90095, USA}

\author{J.~L.~Christiansen}\affiliation{Physics Department, California Polytechnic State University, San Luis Obispo, CA 94307, USA}

\author{A.~J.~Chromey}\affiliation{Center for Astrophysics $|$ Harvard \& Smithsonian, Cambridge, MA 02138, USA}

\author[0000-0003-1716-4119]{A.~Duerr}\affiliation{Department of Physics and Astronomy, University of Utah, Salt Lake City, UT 84112, USA}

\author[0000-0002-1853-863X]{M.~Errando}\affiliation{Department of Physics, Washington University, St. Louis, MO 63130, USA}

\author{M.~Escobar~Godoy}\affiliation{Santa Cruz Institute for Particle Physics and Department of Physics, University of California, Santa Cruz, CA 95064, USA}

\author{S.~Feldman}\affiliation{Department of Physics and Astronomy, University of California, Los Angeles, CA 90095, USA}

\author[0000-0001-6674-4238]{Q.~Feng}\affiliation{Department of Physics and Astronomy, University of Utah, Salt Lake City, UT 84112, USA}

\author[0000-0002-2944-6060]{J.~Foote}\affiliation{Department of Physics and Astronomy and the Bartol Research Institute, University of Delaware, Newark, DE 19716, USA}

\author[0000-0002-1067-8558]{L.~Fortson}\affiliation{School of Physics and Astronomy, University of Minnesota, Minneapolis, MN 55455, USA}

\author[0000-0003-1614-1273]{A.~Furniss}\affiliation{Department of Physics, California State University - East Bay, Hayward, CA 94542, USA}

\author[0000-0002-0109-4737]{W.~Hanlon}\affiliation{Center for Astrophysics $|$ Harvard \& Smithsonian, Cambridge, MA 02138, USA}

\author[0000-0003-3878-1677]{O.~Hervet}\affiliation{Santa Cruz Institute for Particle Physics and Department of Physics, University of California, Santa Cruz, CA 95064, USA}

\author[0000-0001-6951-2299]{C.~E.~Hinrichs}\affiliation{Center for Astrophysics $|$ Harvard \& Smithsonian, Cambridge, MA 02138, USA}\affiliation{Department of Physics and Astronomy, Dartmouth College, 6127 Wilder Laboratory, Hanover, NH 03755 USA}

\author[0000-0002-6833-0474]{J.~Holder}\affiliation{Department of Physics and Astronomy and the Bartol Research Institute, University of Delaware, Newark, DE 19716, USA}

\author[0000-0002-1432-7771]{T.~B.~Humensky}\affiliation{Department of Physics, University of Maryland, College Park, MD, USA}\affiliation{NASA Goddard Space Flight Center, Greenbelt, MD 20771, USA}

\author[0000-0002-1089-1754]{W.~Jin}\affiliation{Department of Physics and Astronomy, University of California, Los Angeles, CA 90095, USA}

\author[0009-0008-2688-0815]{M.~N.~Johnson}\affiliation{Santa Cruz Institute for Particle Physics and Department of Physics, University of California, Santa Cruz, CA 95064, USA}

\author[0000-0002-3638-0637]{P.~Kaaret}\affiliation{Department of Physics and Astronomy, University of Iowa, Van Allen Hall, Iowa City, IA 52242, USA}\affiliation{NASA Marshall Space Flight Center, Huntsville, AL 35812, USA}

\author{M.~Kertzman}\affiliation{Department of Physics and Astronomy, DePauw University, Greencastle, IN 46135-0037, USA}

\author{M.~Kherlakian}\affiliation{DESY, Platanenallee 6, 15738 Zeuthen, Germany}

\author[0000-0003-4785-0101]{D.~Kieda}\affiliation{Department of Physics and Astronomy, University of Utah, Salt Lake City, UT 84112, USA}

\author[0000-0002-4260-9186]{T.~K.~Kleiner}\affiliation{DESY, Platanenallee 6, 15738 Zeuthen, Germany}

\author[0000-0002-4289-7106]{N.~Korzoun}\affiliation{Department of Physics and Astronomy and the Bartol Research Institute, University of Delaware, Newark, DE 19716, USA}

\author[0000-0002-5167-1221]{S.~Kumar}\affiliation{Department of Physics, University of Maryland, College Park, MD, USA }

\author{S.~Kundu}\affiliation{Department of Physics and Astronomy, University of Alabama, Tuscaloosa, AL 35487, USA}

\author[0000-0003-4641-4201]{M.~J.~Lang}\affiliation{School of Natural Sciences, University of Galway, University Road, Galway, H91 TK33, Ireland}

\author[0000-0001-9868-4700]{G.~Maier}\affiliation{DESY, Platanenallee 6, 15738 Zeuthen, Germany}

\author[0000-0003-3802-1619]{M.~Lundy}\affiliation{Physics Department, McGill University, Montreal, QC H3A 2T8, Canada}

\author[0000-0001-7106-8502]{M.~J.~Millard}\affiliation{Department of Physics and Astronomy, University of Iowa, Van Allen Hall, Iowa City, IA 52242, USA}

\author[0000-0001-5937-446X]{C.~L.~Mooney}\affiliation{Department of Physics and Astronomy and the Bartol Research Institute, University of Delaware, Newark, DE 19716, USA}

\author[0000-0002-1499-2667]{P.~Moriarty}\affiliation{School of Natural Sciences, University of Galway, University Road, Galway, H91 TK33, Ireland}

\author[0000-0002-3223-0754]{R.~Mukherjee}\affiliation{Department of Physics and Astronomy, Barnard College, Columbia University, NY 10027, USA}

\author[0000-0002-6121-3443]{W.~Ning}\affiliation{Department of Physics and Astronomy, University of California, Los Angeles, CA 90095, USA}

\author[0000-0002-4837-5253]{R.~A.~Ong}\affiliation{Department of Physics and Astronomy, University of California, Los Angeles, CA 90095, USA}

\author[0000-0001-7861-1707]{M.~Pohl}\affiliation{Institute of Physics and Astronomy, University of Potsdam, 14476 Potsdam-Golm, Germany}\affiliation{DESY, Platanenallee 6, 15738 Zeuthen, Germany}

\author[0000-0002-0529-1973]{E.~Pueschel}\affiliation{Fakult\"at f\"ur Physik \& Astronomie, Ruhr-Universit\"at Bochum, D-44780 Bochum, Germany}

\author[0000-0002-4855-2694]{J.~Quinn}\affiliation{School of Physics, University College Dublin, Belfield, Dublin 4, Ireland}

\author{P.~L.~Rabinowitz}\affiliation{Department of Physics, Washington University, St. Louis, MO 63130, USA}

\author[0000-0002-5351-3323]{K.~Ragan}\affiliation{Physics Department, McGill University, Montreal, QC H3A 2T8, Canada}

\author{P.~T.~Reynolds}\affiliation{Department of Physical Sciences, Munster Technological University, Bishopstown, Cork, T12 P928, Ireland}

\author[0000-0002-7523-7366]{D.~Ribeiro}\affiliation{School of Physics and Astronomy, University of Minnesota, Minneapolis, MN 55455, USA}

\author{E.~Roache}\affiliation{Center for Astrophysics $|$ Harvard \& Smithsonian, Cambridge, MA 02138, USA}

\author[0000-0002-3171-5039]{L.~Saha}\affiliation{Center for Astrophysics $|$ Harvard \& Smithsonian, Cambridge, MA 02138, USA}

\author{M.~Santander}\affiliation{Department of Physics and Astronomy, University of Alabama, Tuscaloosa, AL 35487, USA}

\author{G.~H.~Sembroski}\affiliation{Department of Physics and Astronomy, Purdue University, West Lafayette, IN 47907, USA}

\author[0000-0002-9856-989X]{R.~Shang}\affiliation{Department of Physics and Astronomy, Barnard College, Columbia University, NY 10027, USA}

\author[0000-0002-9852-2469]{D.~Tak}\affiliation{SNU Astronomy Research Center, Seoul National University, Seoul 08826, Republic of Korea}

\author{A.~K.~Talluri}\affiliation{School of Physics and Astronomy, University of Minnesota, Minneapolis, MN 55455, USA}

\author{J.~V.~Tucci}\affiliation{Department of Physics, Indiana University-Purdue University Indianapolis, Indianapolis, Indiana 46202, USA}

\author[0000-0003-2740-9714]{D.~A.~Williams}\affiliation{Santa Cruz Institute for Particle Physics and Department of Physics, University of California, Santa Cruz, CA 95064, USA}

\author[0000-0002-2730-2733]{S.~L.~Wong}\affiliation{Physics Department, McGill University, Montreal, QC H3A 2T8, Canada}

\author[0009-0001-6471-1405]{J.~Woo}\affiliation{Columbia Astrophysics Laboratory, Columbia University, New York, NY 10027, USA}


\correspondingauthor{{M.~J.~Millard}}\email{matthew-j-millard@uiowa.edu}

\begin{abstract}

Observations of GeV gamma-ray emission from the well-studied mixed-morphology supernova remnant (SNR) W44 by Fermi-LAT and AGILE imply that it is a site of significant cosmic ray acceleration. The spectral energy distribution (SED) derived from the GeV data suggest that the gamma-ray emission likely originates from the decay of neutral pions generated by cosmic-ray interactions. It is essential to measure the SED of W44 in the X-ray and very high energy (VHE) gamma-ray bands to verify the hadronic origin of the emission and to gauge the potential contributions from leptonic emission. We report an upper-limit of the nonthermal X-ray flux from W44 of 5 $\times 10^{-13}$ erg cm\textsuperscript{-2} s\textsuperscript{-1} in the 0.5 - 8.0 keV band based on $\sim$ 300 ks of XMM-Newton observations. The X-ray upper limit is consistent with previously estimated hadronic models, but in tension with the leptonic models.  We estimate the VHE flux upper limit of $\sim$ 1.2 $\times$ 10\textsuperscript{-12} erg s\textsuperscript{-1} cm\textsuperscript{-2} in the 0.5 - 5.0 TeV range from W44 using data from the Very Energetic Radiation Imaging Telescope Array System (VERITAS). Our non-detection of W44 at VHE wavlengths is in agreemnent with observations from other imaging atmospheric Cherenkov telescopes (IACTs) and is perhaps consistent with the evolutionary stage of the SNR.

\end{abstract}




\section{Introduction} \label{sec:intro}

 Supernovae (SNe) inject $\sim$ 10\textsuperscript{51} ergs into the interstellar medium (ISM) a few times per century in our Galaxy.   If the remnants of SNe can convert $\sim$ 10\% of the SN explosion energy to cosmic-ray energy, it would suffice to explain the locally observed cosmic-ray population up to the cosmic-ray spectrum knee at 3 PeV \citep{dermer2013}. This energy conversion may be achieved in the expanding shockwaves of supernova remnants (SNRs) through diffusive shock acceleration (DSA), where charged particles gain energy by repeatedly crossing the SNR shock front via scattering off regions of magnetic turbulence \citep{fermi1949,bell1978,reynolds2008}. One of the difficulties in observationally confirming that SNRs are cosmic-ray accelerators is verifying that gamma-ray emission is of hadronic (not leptonic)  origin. Supernova remnant W44 (G34.7-0.4) has one of the strongest lines of evidence that the observed gamma-ray emission results from hadronic cosmic-ray interactions.

 W44  is a prototypical mixed-morphology SNR \citep{rho98}, featuring a well-defined elliptical radio shell \citep{hollinger1966, giacani1997, castelletti2007}, and strong thermal X-ray emission towards the center \citep{harrus1997, rho98, shelton04, uchida2012, okon2020}. Timing measurements of the radio pulsar PSR B1853+01 associated with the remnant suggest a spindown age of $\sim$ 20,000 yr \citep{wolszczan1991, harrus1997}.  W44 has been firmly detected with the high-energy gamma-ray telescopes AGILE \citep{giuliani2011} and the Fermi Large 
 Area Telescope (LAT) \citep{abdo2010}. The best-fit gamma-ray morphology from Fermi-LAT is an elliptical ring with a semi-major axis of 0.3\degr{} that matches well with the radio shell \citep{abdo2010, peron2020}. The spectrum of W44 rises steeply up to $\sim$ 200 MeV, known as the ``pion bump'' \citep{ackermann2013, peron2020}, indicating that the emission originates from the decay of neutral pions produced from energetic proton-proton collisions.   

 Analyses of CO line emission in the vicinity of W44 suggest that the SNR blastwave interacts with dense molecular clouds \citep{sashida2013, seta1998, seta2004, paron2009, yoshiike2013}. Two regions of distinct GeV gamma-ray emission were detected to the southeast and northwest of the SNR \citep{uchiyama2012, peron2020}.  The origin of the emission is unclear, but it may be the result of energetic cosmic rays that escape the shock from W44 and interact with nearby gas clouds or dense ISM.

 W44 is one of the brightest SNRs detected with Fermi-LAT, yet a detection at TeV energies remains elusive. Nonthermal X-ray emission, which may help constrain hadronic emission models, has also not been detected from W44. Here, we present the results of our study of W44 using observations from the Very Energetic Radiation Imaging Telescope Array System (VERITAS), and the X-ray Multi-Mirror Mission (XMM-Newton).  In Section \ref{sec:obs}, we present the details of the observations and data processing.  In Section \ref{sec:anal}, we describe our analyses of the X-ray and very high energy (VHE, $>$ 100 GeV) spectra and results.  In Section \ref{sec:disc} we discuss the results and their implications for the multiwavelength spectral energy distribution (SED).  Section \ref{sec:con} summarizes our findings from our VHE and X-ray analyses of W44.

\section{Observations} \label{sec:obs}

\subsection{VERITAS}
VERITAS is an array of four imaging atmospheric Cherenkov telescopes (IACTs) located at the base of Mt. Hopkins in Amado, AZ (31\degr{} 40\arcmin{} N, 110\degr{} 57\arcmin{} W).  Each of the four telescopes in the VERITAS array has a 12 m diameter optical reflector, consisting of 350 hexagonal spherical mirror segments. The telescopes are spaced, depending on the observing epoch, between 35 m - 170 m apart allowing for stereoscopic observation of the Cherenkov light produced from the shower of secondary particles created from VHE gamma rays interacting with the atmosphere. The mirror placements follow the Davies-Cotton design \citep{davies1957}, for a total reflecting area of 110 m\textsuperscript{2}.  The mirrors focus the Cherenkov light produced from gamma-ray showers onto a camera box at the focal point of the reflector, 12 m from the center of the dish.  The camera consists of 499 Hamamatsu R10560-100-20 26 mm-diameter photomultipler tubes (PMTs), which give a field of view of 3.5\degr{}. The spatial resolution of VERITAS is 0.08\degr{} on the sky at 1 TeV.   The array is capable of detecting a point source with 1\% of the Crab flux after observing it for 25 hours. VERITAS is currently sensitive to photons with energies ranging from 85 GeV to 30 TeV.

We observed W44 with VERITAS in wobble mode for 10 hr and 3.6 hr in May - June of 2008 and 2016, respectively.  For each observing run (38 in total), we pointed the array to a position offset 0.5\degr{}, alternating north, south, east, and west from the target position of ($\alpha_{2000}$ = 284.141\degr, $\delta_{2000}$ = 1.31078\degr{}) in 2008 and ($\alpha_{2000}$ = 284.141\degr{}, $\delta_{2000}$ = 1.32182\degr{}) in 2016  to characterize the detector background rate. We excluded the north wobble data because source-free background measurements were not possible. We applied the standard VERITAS event reconstruction to the remaining data and inspected each run for rate, timing, and tracking consistency. Runs that pass our initial assessment are subjected to further analysis, while rejected runs are not considered. The resulting effective exposure time is 9.6 hr.

\subsection{XMM-Newton}

XMM-Newton observed W44 in 2013 for a total of 335.5 ks. We use data from the three Obs IDs of this observation: 0721630101, 0721630201, and 0721630301.   We focus on the data obtained from the European Photon Imaging Camera (EPIC), which includes two Metal Oxide Semiconductor (MOS) cameras \citep{turner2001} and a pn CCD camera \citep{struder2001}. Due to the poor coverage of the MOS1 camera due to meteroid strikes disabling two of the chips (CCD3 and CCD6) and the higher background count rate of the pn camera, we use data only from the MOS2 camera. We processed the data with the Science Analysis System (SAS) software version 20.0 and the calibration database version 3.13 following the guidelines for extended sources analysis (ESAS) procedures \footnote{https://heasarc.gsfc.nasa.gov/docs/xmm/esas/cookbook/xmm-esas.html}. A total exposure time of 297 ks remained after we performed data quality checks. We generated redistribution matrix files and ancillary response files using the {\tt\textbf{mos-spectra}} command and estimated the non-X-ray background (NXB) using {\tt\textbf{mos-back}}.

\begin{figure}
\plotone{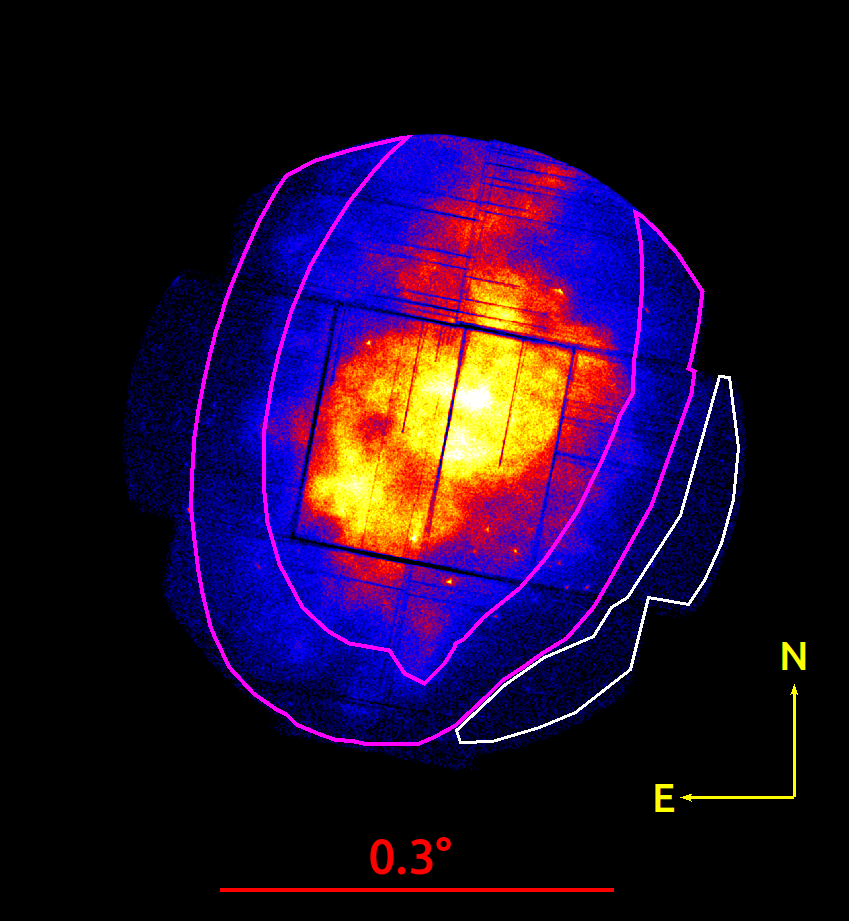}
\caption{XMM MOS2 counts image of supernova
remnant W44 in the 0.5 - 8 keV band smoothed with a gaussian kernel of radius = 3.3\arcsec{}  and $\sigma$ = 1.7\arcsec{}. We extracted the target spectrum from the magenta region and the background spectrum from the white region. 
\label{fig:xmm-image}}
\end{figure}

\hfill 
\break
\hfill
\break

\section{Analysis and Results}  \label{sec:anal}

\subsection{VERITAS}
We processed the data using standard VERITAS analysis techniques, including a Hillas parameter analysis \citep{hillas1985}.  We applied analysis cuts to the mean scaled length, mean scaled width, and integrated charge in the signal to help distinguish between events of cosmic-ray and gamma-ray origins. We performed the event reconstruction with the Eventdisplay software \citep{maier2017} and converted the data to DL3 fits files compatible with the Gammapy (v1.1) \citep{gammapy2023, aguasca_cabot2023} software for further analysis. Our spectral energy threshold is 0.4 TeV. We extracted the spectrum from a circle of radius, $\theta$ = 0.3\degr{}, centered on the approximate radio-image position of W44 ($\alpha_{2000}$ = 284.015\degr{}, $\delta_{2000}$ = 1.349\degr{}). For the background spectrum, we selected between 2 and 3 locations within the field of view according to the reflected regions method \citep[see][]{berge2007}. This method does not rely on accounting for the radial dependence of the background acceptance. The statistical significance of any excess is evaluated using the likelihood method described by \citet{li+ma1983}, which takes into account the counts from the on-source and background regions and their different acceptances.

We find no significant source emission at the location of W44.  We estimate a differential flux upper limit of $\sim$ 1.2 $\times$ 10\textsuperscript{-12} erg s\textsuperscript{-1} cm\textsuperscript{-2} in the 0.5 - 5.0 TeV range, for a powerlaw model with a spectral index of 2.4.  The upper limit in each energy bin corresponds to a 95\% confidence level using the Wstat\footnote{https://docs.gammapy.org/dev/api/gammapy.\\stats.wstat.html\#gammapy.stats.wstat} test-statistic value. Our flux upper limits agree well with a powerlaw extrapolation of the Fermi-LAT data from \citet{peron2020} and are somewhat more constraining than the H.E.S.S. upper limit in the 1 - 10 TeV band \citep{hess2018}.

\begin{figure*}
\plotone{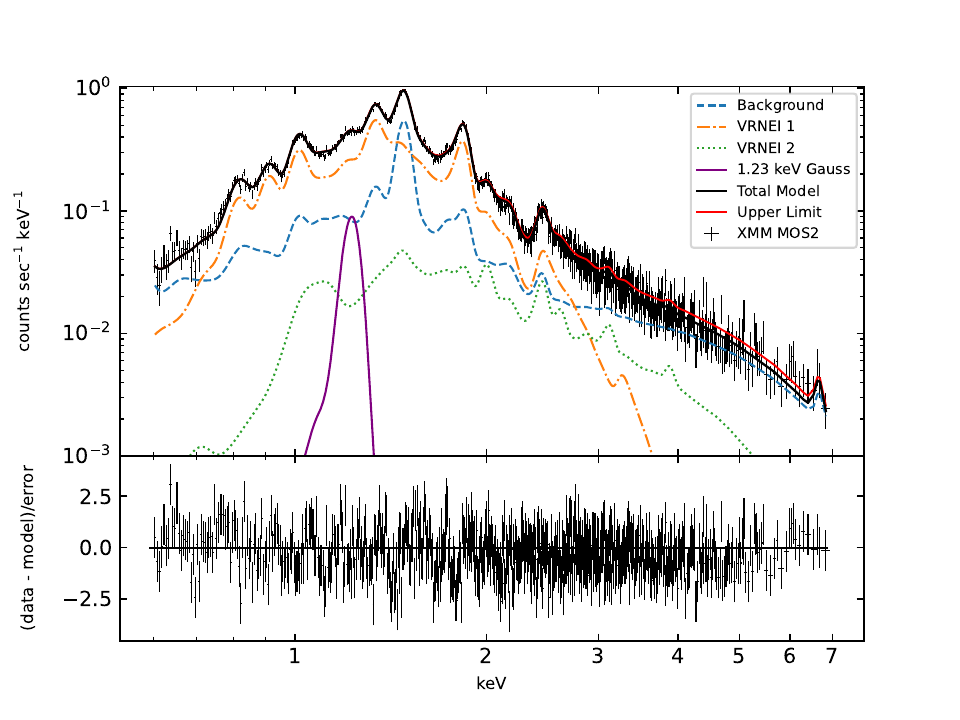}
\caption{XMM-Newton MOS2 spectrum extracted from the radio boundary of SNR W44 (magenta region in Figure \ref{fig:xmm-image}) with our best-fit model overlaid (solid black curve). The solid red curve shows the best-fit model with the nonthermal flux upper limit included.
\label{fig:xmm-spec}}
\end{figure*}

%


\subsection{XMM-Newton}
We extracted the XMM MOS2 spectrum from a region coinciding with the observed GeV gamma-ray ring-like emission from W44 (see the magenta region in Figure \ref{fig:xmm-image}). We ﬁtted the 0.5 – 8 keV spectrum with an absorbed X-ray emission spectral model assuming two variable recombining non-equilibrium ionization plasma models, {\tt\textbf{tbabs}} * ({\tt\textbf{vrnei}} + {\tt\textbf{vrnei}}), using the XSPEC software package \citep{arnaud96}. We also included a gaussian line at $\sim$ 1.25 keV to account for unknown excess emission possibly due to missing Fe L lines in the atomic data \citep{brickhouse2000}. Figure \ref{fig:xmm-spec} shows the extracted spectrum and our best-fit emission + background model. We subtracted the NXB from the spectrum and estimated the non-NXB background emission by modeling the spectrum extracted from the white region in Figure \ref{fig:xmm-image}. Our background model includes the following components: foreground emission, Galactic ridge X-ray emission, the cosmic X-ray background, and Al and Si K$\alpha$ instrumental lines.   

Our best-fit {\tt\textbf{vrnei}} model parameters are summarized in Table \ref{tab:vrnei}. In our lower-kT {\tt\textbf{vrnei}} model, the Ne, Mg, Si, S, and Fe abundances are allowed to vary.  The Ni abundance is tied to the Fe abundance, while the Ar and Ca abundances are tied to the S abundance. In the higher-kT {\tt\textbf{vrnei}} model, the abundances are fixed at solar values \citep{wilms2000}. The two absorbed {\tt\textbf{vrnei}} components can reasonably account for the observed source emission in the X-ray spectrum of W44 with $\chi^2$/dof = 684/716.  

To estimate an upper limit on the nonthermal emission from W44, we added a {\tt\textbf{powerlaw}} component to our model. We fixed the photon index at the standard value for SNRs, $\Gamma$ = 2. Using the {\tt\textbf{steppar}} command in XSPEC, we increased the {\tt\textbf{powerlaw}} flux until the $\chi^2$ statistic reached the value corresponding to the 99\% confidence level above the minimum,  giving a flux value of 2.8 $\times$ 10\textsuperscript{-13} ergs cm\textsuperscript{-2} s\textsuperscript{-1}. Although the shape of our extraction region is based on the best-fit ring found by \citet{peron2020}, it only covers about 60\% of the ring area due to the constraint of the XMM MOS field of view. To account for the lack of coverage, we multiply our estimated upper limit by a factor of 1.7. Thus, we estimate a flux upper limit $\sim$ 5 $\times$ 10\textsuperscript{-13} ergs cm\textsuperscript{-2} s\textsuperscript{-1} in the 0.5 - 8.0 keV band. 

To test the effect of the photon index on the upper limit, we tried several values, from $\Gamma$ = 1.2 to 2.8. For $\Gamma$ $\gtrsim 2.4$ and $\Gamma$ $\lesssim 1.6$, the resulting change in upper limit is enough to affect our conclusions regarding the preferred multiwavelength emission model (discussed in Section \ref{sec:disc}). However, we note that the measured photon indices of other mixed-morphology SNRs with potential nonthermal emission, W28 and G346.6-0.2, are $\Gamma \sim$ 2 \citep{zhou2014, auchettl2017}. Moreoever, higher photon indices ($\Gamma$ $\gtrsim 2.5$) are generally expected only for younger SNRs, less than a few thousand years old. Thus, our choice of $\Gamma$ = 2 is suitable for W44, a $\sim 2 \times 10^4$ yr old mixed-morphpology SNR.

\begin{deluxetable}{llc}
\tablecaption{Best-fit X-ray Spectral Model Parameters \label{tab:vrnei}}
\tablewidth{0pt}
\tablehead{\colhead{Component} & \colhead{Parameter} & \colhead{}}
\startdata
  $N_H$ (cm\textsuperscript{-2}) &  & 2.32$\pm 0.10\times 10^{21}$  \\ 
  vrnei  & $kT_e$ (keV) & 0.215 $^{+0.004}_{-0.007}$ \\ 
  & $n_et$ (cm\textsuperscript{-3} s) & 8.28 $^{+0.81}_{-0.76}\times 10^{11}$ \\ 
  & Norm & 0.12 $\pm$ 0.04 \\
  & Ne &  4.1 $^{+3.0}_{-1.5}$\\ 
  & Mg & 2.8 $^{+1.7}_{-0.9}$ \\ 
  & Si & 5.6 $^{+13.9}_{-1.9}$ \\
  & S &  5.8 $^{+4.8}_{-2.6}$\\
  & Fe &  4.6 $^{+4.9}_{-2.1}$\\
  \hline
  vrnei & $kT_e$ (keV) & 1.70 $^{+0.66}_{-0.34}$ \\ 
  & $n_et$ (cm\textsuperscript{-3} s) & 2.12 $^{+6.74}_{-1.34}\times 10^{11}$ \\ 
  & Norm & 0.003 $\pm$ 0.001
\enddata
\tablecomments{Elemental abundances are relative to solar values \citep{wilms2000}.}
\end{deluxetable}


%


\begin{figure*}
\plotone{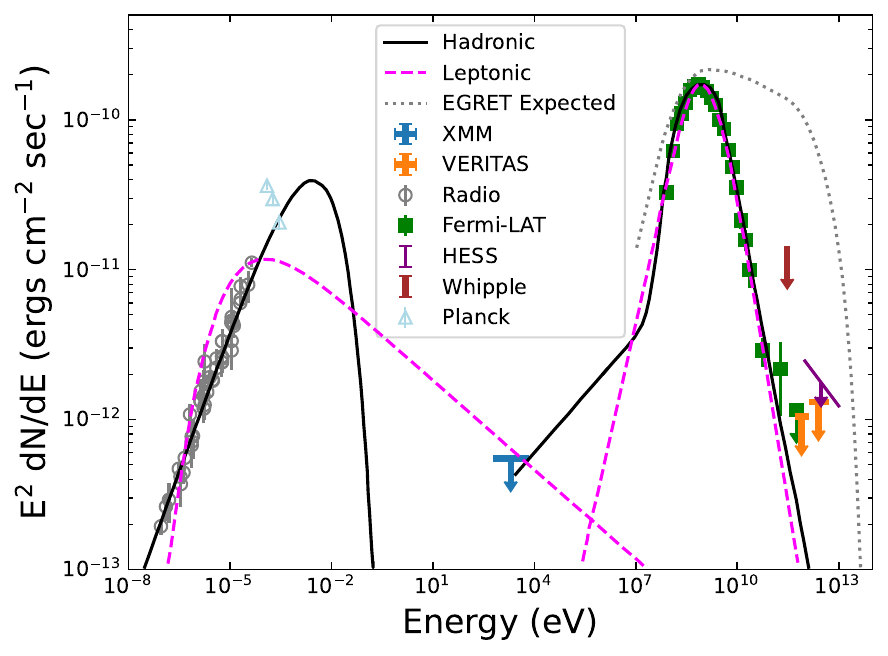}
\caption{Multiwavelength SED of W44. Radio data are from \citet{castelletti2007}, Planck data from \citet{planck2016}, Fermi-LAT data from \citet{peron2020}, and HESS upper limits from \citet{hess2018}.  The overlaid best-fit hadronic (solid black lines) and leptonic model (dashed magenta lines) are adapted from Figures 4 and 5 of \citet{cardillo2014} and correspond to models `H3' and `L2', respectively, in their Table 3. The hadronic model includes a bremsstrahlung component, which dominates from $\sim$ 10\textsuperscript{3} - 10\textsuperscript{7} eV The grey dotted line denotes the expected emission model from EGRET observations adapted from Figure 8 of \citet{buckley1998}.  The XMM-Newton and VERITAS upper limits are from this work.
\label{fig:w44-sed}}
\end{figure*}

%


\section{Discussion}  \label{sec:disc}

\subsection{Multiwavelength SED}

\citet{giuliani2011} modeled the SED of W44 using radio \citep{castelletti2007}, AGILE, and Fermi-LAT data. The authors tested both hadronic- and leptonic-based models and found that a hadronic model fit best for cosmic rays interacting with a gas of density $n_o$ = 100 cm\textsuperscript{-3}. \citet{cardillo2014} analyzed a newer AGILE dataset together with CO data from the NANTEN2 telescope and found a similar result. The best-fit hadronic (a broken powerlaw with $p_1$ = 2.2, $p_2$ = 3.2, and $E_{break}$ = 20 GeV) and leptonic (smoothed broken powerlaws, with $p'_1$ = -2.5, $p'_2$ = 3.4, and $E_{break}$ = 0.5 GeV) SED models from \citet{cardillo2014} are shown in Figure \ref{fig:w44-sed}. The leptonic models are dominated by bremsstrahlung emission at gamma-ray wavelengths. The hadronic model also includes a weak bremsstrahlung component. \citet{peron2020} similarly found that a hadronic model was the best fit for the gamma-ray SED based on 9.7 years of Fermi-LAT data. Thus, model fitting of the SED of W44 strongly suggests that the emission is hadronic in origin, likely from cosmic rays accelerated by the SNR blast wave encountering dense gas clouds, interacting via proton-proton reactions, and producing neutral pions that decay via gamma-ray emission. Confirmation of the hadronic emission scenario with the multiwavelength SED is essential, since a pion bump feature can also originate from diffuse gamma-ray emission of the Galactic plane \citep{hunter1997}.  Our non-thermal X-ray flux upper limit is plotted along with the best-fit hadronic model from \citet{cardillo2014} in Figure \ref{fig:w44-sed}. Our upper limit slightly favors the best-fit hadronic models over the leptonic models from \citet{cardillo2014}.   

Initial gamma-ray observations of W44 with the Energetic Gamma-Ray Experiment Telescope (EGRET) on board the Compton Gamma Ray Observatory found flux levels potentially consistent with cosmic-ray acceleration up to PeV energies \citep{esposito1996}. The model flux from \citet{buckley1998} estimated from EGRET measurements and extrapolated to PeV energies is shown by the grey dotted curve in Figure \ref{fig:w44-sed}. However, flux upper limits estimated from follow-up observations with the Whipple telescope were well below the expected TeV flux level \citep{buckley1998}.   Although W44 is one of the brightest SNRs observed at GeV energies \citep{acero2016}, the flux decreases with a powerlaw index of $3.3 \pm 0.4$ above 10 GeV \citep{peron2020}, suggesting that W44 should be relatively faint at VHE wavelengths.

Our VERITAS flux upper limit does not constrain the predicted emission models of \citet{cardillo2014} and \citet{peron2020}, which fall off steeply above 10 GeV. Observations of the W44 region with significantly deeper exposures using the High Energy Stereoscopic System (HESS) and the Major Atmospheric Gamma Imaging Cherenkov (MAGIC) telescopes have also not yielded a detection of W44 \citep{hess2018,ditria2022} at VHE wavelengths. \citet{brose2020} modeled the evolution of the cosmic ray spectrum inside a supernova remnant up to an age of 10\textsuperscript{5} yr. The leptonic emission declines more rapidly than the hadronic emission due to energy losses of electrons. However, after 10\textsuperscript{4} yr, the hadronic flux would also be low, and thus older remnants can be seen only when the gas density is high. Hence, the interaction of old SNRs with massive molecular clouds may be needed to produce significant VHE gamma-ray emission. \citet{brose2021} produced simulated gamma-ray SEDs for test SNRs assuming a distance of 1 kpc. After 10\textsuperscript{4} yr, the SED appears soft, owing to a rapid decline in the maximum energy to which the shock can currently accelerate particles.  Considering the estimated age of W44 of $\sim$ 20 kyr, this could be the case in W44 as well.

IC 443 is another SNR that shows evidence of significant spectral curvature in the GeV band and a hadronic origin for the gamma-ray emission \citep{tavani2010}. Interestingly, W44 and IC 443 are both mixed-morphology SNRs on the order of 10\textsuperscript{4} years old.  However, IC 443 has been firmly detected in the VHE range \citep{albert2007,acciari2009}.  The difference in VHE flux levels is perhaps due to the age of the SNRs. The age estimates for IC 443 vary significantly, from 3 kyr \citep{petre1988} to 30 kyr \citep{chevalier1999,olbert2001}. Thus, IC 443 may be in a younger evolutionary stage than W44.

%


\subsection{Thermal X-ray Emission at the SNR Boundary}

The hadronic models for gamma-ray emission strongly depend on the assumed ambient density around the cosmic-ray source.  Both high- and low-density scenarios have been considered by \citet{cardillo2014} and \citet{peron2020}, respectively, when modeling the hadronic gamma-ray flux.  Both studies use a broken powerlaw function to model the proton spectrum. However, \citet{cardillo2014} found a best-fit ambient density of 300 cm\textsuperscript{-3} for a distance, $d$ = 3.1 kpc \citep{clark1976,wolszczan1991}, while \citet{peron2020}) found $n_o$ = 10 cm\textsuperscript{-3} for $d$ = 2.2 kpc. Although their best-fit ambient densities and assumed distances are different, \citet{cardillo2014} and \citet{peron2020} report a similar total proton energy in the remnant, $W_p$, with values of $5 \times 10^{49}$ erg and $1.2 \times 10^{49}$ erg, respectively, with both models fitting the gamma-ray SED reasonably well. This is possible because $W_p$ is proportional to the square of the distance and inversely proportional to density.   

We can use our model of the thermal X-ray emission at the boundary of W44 to estimate the ambient density around the expanding SNR by applying the equation for ram pressure,  $P_{ram}$ $\thickapprox$ $\rho_o(V_s)^2$, where $V_s$ is the shock speed, and $\rho_o$ is the ambient mass density.  The ion pressure interior to the shock boundary should be equal to the ram pressure, $P_{ram} = P_{ion} \thickapprox n_{ion}(kT_{ion})$, where $n_{ion}$ and $kT_{ion}$ are the ion number density and temperature, respectively.  Thus, we can solve for the ambient number density, $n_o = \rho_o/m_H = n_{ion}(kT_{ion})/(m_HV_s^2)$, where $m_H$ is the mass of hydrogen.

The lower-kT {\tt\textbf{vrnei}} component dominates our model, with temperature $kT_e$ $\sim$ 0.2 keV and ionization timescale $n_et$ $\sim$ 8 $\times$ 10\textsuperscript{11} s cm\textsuperscript{-3}. The high ionization timescale (nearly 10\textsuperscript{12} s cm\textsuperscript{-3}) suggests that W44 is approaching collisional ionization equilibrium. Thus, the electrons and ions would have similar temperatures, \textit{ \textit{kT\textsubscript{e}}  $\thickapprox$ kT\textsubscript{ion}} = 0.2 keV. The electron density, $n_e$, may be determined if the time since the plasma was shocked, $t$, is known. Using a conservative \textit{t} of 10\textsuperscript{4} years gives an electron number density, $n_e$ $\sim$ 2 cm\textsuperscript{-3}. However,  based on CO line emission, \citet{anderl2014} estimated the age of shocks in the northeast of the remnant to be as young as $\sim$ 10\textsuperscript{3} years, which would imply $n_e$ $\sim$ 20 cm\textsuperscript{-3}.   Assuming solar elemental abundances for the emitting plasma, $n_e$ = 1.2$n_H$, where $n_H$ is the number density of hydrogen in the plasma. Solving for $n_H$, and given that $n_H \thickapprox n_{ion}$, then $n_{ion} = 1.67 - 16.7$ cm\textsuperscript{-3}, which implies $P_{ram}$ = $5.4 \times 10^{-10} - 54 \times 10^{-10}$ dyn cm\textsuperscript{-2}. 

\citet{Koo1995} found that the shell of H I gas around W44 has an expansion speed, $V_s$ = 150 km s\textsuperscript{-1}.  This is generally taken to be the forward shock speed for W44 \citep{cox1999,tang2015}. However, inserting this value into our equation for ram pressure gives a post-shock to pre-shock density (compression) ratio of approximately unity, instead of the factor of 4 expected for a strong shock in adiabatic conditions. We note that \citet{Koo1995} suggested that the H I shell was separate from the radio continuum shell, for which they estimated an expansion speed of 330 km s\textsuperscript{-1}. \citet{shelton1999} also estimated shock speeds up to $\sim 300$ km s\textsuperscript{-1} for the tenuous region of W44. Thus, we assume $V_s$ = 300 km s\textsuperscript{-1}, which gives a compression ratio, $n_{ion}/n_o = 4.6$, much closer to the expected ratio. We estimate the ambient density to be between 0.36 and 3.6 cm\textsuperscript{-3}.  Our value is in agreement with \citet{cox1999}, who found an ambient density of about 6 cm\textsuperscript{-3} using ROSAT data, assuming that the shell formation time equals the age. Thus, the density estimated from our X-ray measurements is more consistent with the scenario of a closer distance of $\sim$ 2 kpc to W44. 

We note that CO line measurements of molecular gas around W44 \citep{yoshiike2013} suggest that the molecular proton density is about 200 cm\textsuperscript{-3} on average. The submillimeter density estimate includes some high-density knots, which could dominate the average causing it to be significantly higher than the X-ray estimate.  For such a high average density, the Sedov radius and post-shock temperature would be too small compared to what we observe. In this scenario, the shock likely initially passed through low-density material in a wind-blown bubble created by the SN progenitor, where the shock was fast \citep{das2024}. The forward shock would eventually interact with the dense gas, slow down until $T_e$ is too small for X-ray emission, and effectively produce gamma rays in the high-density environment. \citet{das2024} found that the spectral index for pion-decay emission would exhibit a soft proton spectra, reaching 2.4 - 2.6 above 10 GeV, similar to the soft gamma-ray spectrum observed in W44.


\section{Conclusions}  \label{sec:con}

We investigated the X-ray and VHE emission from SNR W44 using data from the XMM-Newton and VERITAS telescopes to better understand the origin of the multiwavelength emission. Our analysis of the X-ray spectrum at the boundary of the SNR did not reveal the presence of nonthermal emission coincident with observed GeV emission. However, we note that our nonthermal X-ray flux upper estimate of 5 $\times 10^{-13}$ erg cm\textsuperscript{-2} s\textsuperscript{-1} is consistent with the hadronic model and is in tension with the leptonic model from \citet{cardillo2014}.  We used our best-fit model of the thermal X-ray emitting plasma to estimate the ambient density around the SNR. Our density estimate is generally consistent with other X-ray analyses, and may favor a distance of $\sim$ 2 kpc to W44 when modeling the gamma-ray SED assuming a hadronic origin. Based on its bright GeV emission, it was perhaps expected that W44 might be a contributor of Galactic PeV cosmic rays. However, our analysis of the VERITAS observations of W44 also did not result in a detection.  Our flux upper limit slightly improves upon the Fermi-LAT and HESS upper limits in the TeV band, and confirms that W44 is significantly fainter in the TeV compared to the GeV band. Future gamma-ray observatories, like the Cherenkov Telescope Telescope Array Observatory (CTAO), may be able to detect W44 in the VHE regime.  The expected flux sensitivity of the CTAO based on a 50-hour exposure is similar to the expected flux from W44 from 0.1 - 1.0 TeV \citep{ctao2019}. Thus, an exposure of this duration could constrain the current best-fit emission models.  W44 is a exemplar case of a sedov phase SNR emitting in gamma-rays via hadronic interactions. The presence of a cutoff below TeV energies points to questions about the acceleration mechanism and propagation models that suggest SNRs are able to produce cosmic rays with energies up to the knee of the locally observed cosmic ray spectrum.

\begin{acknowledgments}
This research is supported by grants from the U.S. Department of Energy Office of Science, the U.S. National Science Foundation and the Smithsonian Institution, by NSERC in Canada, and by the Helmholtz Association in Germany. This research used resources provided by the Open Science Grid, which is supported by the National Science Foundation and the U.S. Department of Energy's Office of Science, and resources of the National Energy Research Scientific Computing Center (NERSC), a U.S. Department of Energy Office of Science User Facility operated under Contract No. DE-AC02-05CH11231. We acknowledge the excellent work of the technical support staff at the Fred Lawrence Whipple Observatory and at the collaborating institutions in the construction and operation of the instrument. M. M. thanks NSF for support under grant PHY-2012916. We are grateful for the suggestions provided by the anonymous referee, which helped us to improve the clarity and quality of this work.
\end{acknowledgments}

%

\vspace{5mm}
\facilities{VERITAS, XMM-Newton}


\software{astropy \citep{2013A&A...558A..33A,2018AJ....156..123A}, Eventdisplay \citep{maier2017}, 
          Gammapy \citep{gammapy2023}, XSPEC \citep{arnaud96}
          }





\bibliographystyle{aasjournal}



\end{document}